\documentclass{article}
\usepackage{spconf,amsmath,graphicx,hyperref}
\usepackage{booktabs}
\usepackage{amsmath}
\usepackage{amssymb}
\usepackage{multirow}
\usepackage{xcolor}
\usepackage{color, colortbl}
\usepackage{kotex}
\usepackage{lipsum}
\usepackage{tabularx}

\title{MULTIMODAL SELF-ATTENTION NETWORK WITH TEMPORAL ALIGNMENT\\ FOR AUDIO-VISUAL EMOTION RECOGNITION}
\name{Inyong Koo$^{1}$ \qquad Yeeun Seong$^{2}$ \qquad Minseok Son$^{1}$ \qquad Jaehyuk Jang$^{1}$ \qquad Changick Kim$^{1}$}
\address{
  Korea Advanced Institute of Science and Technology (KAIST)\\
  $^{1}$School of Electrical Engineering \quad
  $^{2}$Graduate School of Green Growth and Sustainability\\
  Daejeon, South Korea
}
\begin{document}
%\ninept
%
\maketitle
\begin{abstract}
%Existing audio-visual emotion recognition (AVER) methods typically extract utterance-level features from each modality and then fuse them for classification. 
%Although attention-based approaches have considered frame-level dependencies across modalities, the frame-rate mismatch between audio and video features remains largely unaddressed.
Audio-visual emotion recognition (AVER) methods typically fuse utterance-level features, and even frame-level attention models seldom address the frame-rate mismatch across modalities.
In this paper, we propose a Transformer-based framework focusing on the temporal alignment of multimodal features.
Our design employs a multimodal self-attention encoder that simultaneously captures intra- and inter-modal dependencies within a shared feature space.
To address heterogeneous sampling rates, we incorporate Temporally-aligned Rotary Position Embeddings (TaRoPE), which implicitly synchronize audio and video tokens.
Furthermore, we introduce a Cross-Temporal Matching (CTM) loss that enforces consistency among temporally proximate pairs, guiding the encoder toward better alignment.
Experiments on CREMA-D and RAVDESS datasets demonstrate consistent improvements over recent baselines, suggesting that explicitly addressing frame-rate mismatch helps preserve temporal cues and enhances cross-modal fusion.
\end{abstract}
\begin{keywords}
Multimodal emotion recognition, Transformer encoder, temporal alignment, Rotary Position Embedding (RoPE), cross-temporal matching loss
\end{keywords}

\section{INTRODUCTION}
\label{sec:intro}

Understanding human emotion from multimodal signals such as speech and facial expressions is a central challenge in affective computing and human–computer interaction~\cite{ survey_lian}.
Audio-visual emotion recognition (AVER) benefits from complementary cues across modalities: prosody and intonation in speech often co-occur with facial action units and expressions in video.
However, effectively modeling and synchronizing these heterogeneous signals remains a long-standing problem.

Early AVER systems typically extract modality-specific representations at the utterance level, and then fuse them with simple concatenation for classification~\cite{luna2021multimodal, lei2023audio, taavn_radoi, luna2021proposal, middya2022deep}.
% ~\cite{vaanet_zhao, commoninfo_ma, guo_conformer}.
This design is simple and robust, but collapses modality-specific temporal dynamics into a global embedding, limiting fine-grained cross-modal interactions.
More recent works attempt to exploit cross-modal dependencies between audio and video sequences using a variety of cross-attention architectures~\cite{attanet_fan, mocanu2023multimodal, sun2024hicmae, sadok2025vector}, yet they typically rely only on feature-level similarity and do not incorporate temporal cues.
% ~\cite{cross_praveen, multimodal_mocanu, rcma_praveen, attanet_fan}
In particular, Transformer-based architectures \cite{transformer_vaswani, attsfnet_zhang} capture ordering within each modality through positional embeddings, but remain agnostic to the relative temporal structure across modalities, leaving cross-modal synchronization insufficiently modeled.

This challenge becomes evident when we consider the different temporal granularities of audio and video representations.
Audio features are extracted at a finer temporal resolution (e.g., 50 FPS), whereas video features are typically coarser (e.g., 30 FPS), which leads to asynchronous token sequences.
In the absence of synchronization, cross-modal attention can scatter over irrelevant positions, weakening fine-grained multimodal associations.

In this paper, we focus on multimodal fusion from the perspective of temporal alignment.
We propose a Transformer-based framework that projects audio and video features into a shared embedding space, where a unified multimodal self-attention encoder simultaneously captures intra- and inter-modal dependencies.
To resolve the frame-rate disparity between streams, we incorporate Temporally-aligned Rotary Position Embeddings (TaRoPE),  a variant of RoPE~\cite{rope_su} designed to implicitly synchronize tokens along a consistent temporal axis.
In addition, we introduce a Cross-Temporal Matching (CTM) loss that leverages timestamp-aware Gaussian affinities to encourage similarity among temporally proximate audio-visual pairs.

We validate our approach on two benchmark datasets, CREMA-D~\cite{crema-d_cao} and RAVDESS~\cite{ravdess_livingstone}. 
Experimental results show that our method outperforms strong attention-based baselines, demonstrating that explicitly modeling frame-rate mismatch achieves more effective multimodal fusion.
\begin{figure*}[th]
\centering
  \includegraphics[width=0.95\linewidth]{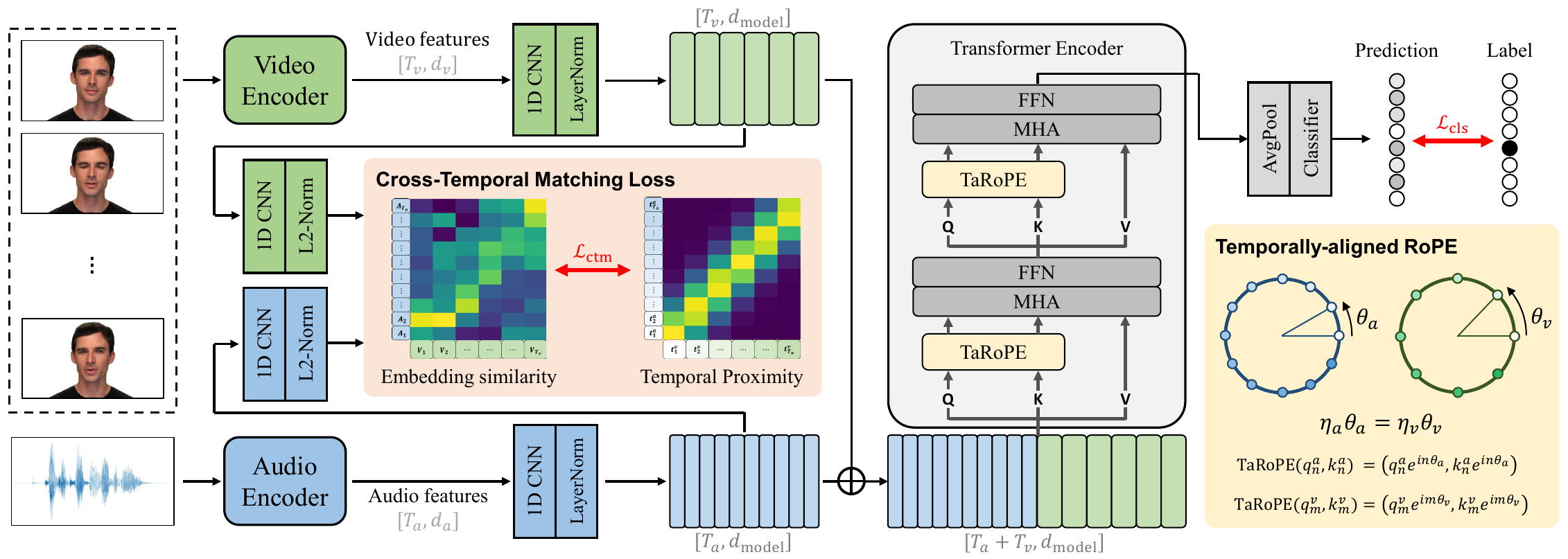}
  \caption{Audio and video features are first extracted by modality-specific encoders and projected into a shared embedding space. Temporally-aligned RoPE (TaRoPE) implicitly synchronizes heterogeneous audio–visual sequences within the unified Transformer encoder by adapting rotary positional embeddings to different temporal resolutions.  The Cross-Temporal Matching (CTM) loss explicitly enforces temporal consistency by aligning audio–video feature pairs that are temporally proximal along a shared time axis. The final representation is pooled for emotion classification.}
  \label{fig:overview}
\end{figure*}

\section{METHODOLOGY}
\label{sec:methods}

This section describes the architecture for extracting and fusing audio-visual features, Temporally-aligned RoPE (TaRoPE) for implicit synchronization, and the Cross-Temporal Matching (CTM) loss for explicit temporal consistency.
Figure~\ref{fig:overview} presents an overview of the proposed framework.

\subsection{Architecture}

The framework first processes raw audio and video inputs with dedicated encoders to extract modality-specific features.
For the audio modality, we employ a pretrained xlsr-Wav2Vec 2.0 encoder~\cite{xlsr_conneau}, which produces 1024-dimensional frame-level embeddings.
Given an input sampled at 16kHz, the encoder’s internal downsampling factor of 320 results in an effective frame rate of $\eta_a =50$ FPS.
For the video modality, we extract 35 Action Unit feature descriptors, which are muscle-based features defined by the Facial Action Coding System (FACS) \cite{facs_friesen}, using the OpenFace library~\cite{openface_baltruvsaitis} from each face frame at $\eta_v = 30$ FPS.

To facilitate multimodal integration, both audio and video features are first linearly projected into a shared $d_{\text{model}}$-dimensional embedding space:
\begin{align}
\textbf{F}^a = [f_1^a, f_2^a, \ldots, f_{T_a}^a] \in \mathbb{R}^{T_a \times d_{\text{model}}}, \\[4pt]
\textbf{F}^v = [f_1^v, f_2^v, \ldots, f_{T_v}^v] \in \mathbb{R}^{T_v \times d_{\text{model}}},
\end{align}
where $\textbf{F}^a$ and $\textbf{F}^v$ are sequences of frame-level audio and video features of lengths $T_a$ and $T_v$, respectively.
These sequences serve as the token embeddings fed into the unified Transformer encoder, which incorporates two multimodal self-attention blocks with TaRoPE to ensure temporal alignment during attention.
Temporal average pooling is then applied to obtain an utterance-level representation.
Finally, this representation is passed to a softmax classifier to predict the target emotion.

\subsection{Temporally-aligned RoPE}

A fundamental yet underexplored issue in multimodal fusion is the mismatch in temporal resolution between audio and video sequences.
To address this, we propose Temporally-aligned Rotary Position Embeddings (TaRoPE), which adapt RoPE~\cite{rope_su} to synchronize heterogeneous frame rates.

In the standard RoPE formulation, positional information is injected into the query and key vectors $q_n, k_n \in \mathbb{R}^{d_{\text{model}}}$ (projected from the token embedding $x_n$) by rotating them according to their indices:
\begin{equation}
(q_n, k_n) \mapsto (q_n e^{i n \theta_{\text{base}}}, \; k_n e^{i n \theta_{\text{base}}}),
\end{equation}
where $n$ denotes token positions and $\theta_{\text{base}}$ is the base frequency. 
This construction ensures that the attention score between positions $n$ and $m$, 
$\langle \text{RoPE}(q_n), \text{RoPE}(k_m)\rangle$, depends only on their index distance $(n-m)$, 
thereby encoding relative positional information directly in the self-attention mechanism.

Let $f^a_n$ and $f^v_m$ denote the features of $n$-th audio frame and the $m$-th video frame, 
from which query and key vectors are derived. 
TaRoPE applies modality-specific rotations as
\begin{align}
(q^a_n, k^a_n) &\mapsto (q^a_n e^{i n \theta_a}, \; k^a_n e^{i n \theta_a}), \\[4pt]
(q^v_m, k^v_m) &\mapsto (q^v_m e^{i m \theta_v}, \; k^v_m e^{i m \theta_v}),
\end{align}
with $\theta_v = \tfrac{\eta_a}{\eta_v}\theta_a$. 
This rescales video positions to the audio timeline, 
ensuring that cross-modal attention depends on consistent temporal distances.

\subsection{Cross-Temporal Matching Loss}
To explicitly enforce temporal consistency, we introduce a Cross-Temporal Matching (CTM) loss that encourages audio and video frames that are temporally proximal along a shared physical time axis to share similar representations.

Before the Transformer encoder, both modalities are projected into an L2-normalized embedding space $\mathbb{R}^{d_{\text{emb}}}$. 
Given embeddings $\textbf{E}^a = [A_1, A_2, \ldots, A_{T_a}]$ and $\textbf{E}^v = [V_1, V_2, \ldots, V_{T_v}]$ with timestamps $t_i^a$ and $t_j^v$ defined on a common timeline, we define a temporal Gaussian affinity as:
\begin{equation}
g_{ij}=\exp\!\Big(-\tfrac{(t^a_i-t^v_j)^2}{2\sigma^2}\Big),
\end{equation}
where $\sigma$ controls the tolerance to misalignment. 
Feature similarity is measured as $s_{ij}=A_i^\top V_j$, which is normalized via a temperature-scaled softmax ($\tau>0$). 
The predicted distributions $p$ are thus obtained from $\{s_{ij}\}$, while the target distributions $q$ are derived from $\{g_{ij}\}$: 
\begin{align}
p^{a\to v}_{ij} &= \frac{\exp(s_{ij}/\tau)}{\sum_{j'} \exp(s_{ij'}/\tau)}, &
q^{a\to v}_{ij} &= \frac{g_{ij}}{\sum_{j'} g_{ij'}}, \\
p^{v\to a}_{ij} &= \frac{\exp(s_{ij}/\tau)}{\sum_{i'} \exp(s_{i'j}/\tau)}, &
q^{v\to a}_{ij} &= \frac{g_{ij}}{\sum_{i'} g_{i'j}}.
\end{align}

The CTM loss aligns these distributions using bidirectional cross-entropy:
\begin{align}
\mathcal{L}_{a\to v} &= \tfrac{1}{T_a}\sum_{i=1}^{T_a}
\mathrm{CrossEntropy}\!\big(q^{a\to v}_{i\cdot},\,p^{a\to v}_{i\cdot}\big), \\[4pt]
\mathcal{L}_{v\to a} &= \tfrac{1}{T_v}\sum_{j=1}^{T_v}
\mathrm{CrossEntropy}\!\big(q^{v\to a}_{\cdot j},\,p^{v\to a}_{\cdot j}\big), \\[4pt]
\mathcal{L}_{\text{ctm}} &= \tfrac{1}{2}\big(\mathcal{L}_{a\to v}+\mathcal{L}_{v\to a}\big).
\end{align}
Here, the notation $q^{a\to v}_{i\cdot}$ and $p^{a\to v}_{i\cdot}$ refer to the $i$-th row distributions over all video frames, while $q^{v\to a}_{\cdot j}$ and $p^{v\to a}_{\cdot j}$ denote the $j$-th column distributions over all audio frames.

The final training objective combines this with the classification loss:
\begin{equation}
\mathcal{L}_{\text{total}}
= \mathcal{L}_{\text{cls}} + \lambda_{\text{ctm}}\,\mathcal{L}_{\text{ctm}},
\end{equation}
where $\lambda_{\text{ctm}}$ denotes the weight for CTM loss.

\section{EXPERIMENTS}
\label{sec:experiments}

\subsection{Datasets and Evaluation Metrics}
\label{ssec:datasets}

The CREMA-D dataset~\cite{crema-d_cao} contains 7,442 clips of short sentences spoken by 91 actors (48 female, 43 male) with diverse age and ethnic backgrounds.
Each utterance is labeled with one of six basic emotions: anger, disgust, fear, happy, neutral, and sad.
For evaluation, we adopt the speaker-independent train/validation split provided by \cite{emobox_ma}.

The RAVDESS dataset \cite{ravdess_livingstone} consists of 1,440 utterances produced by 24 professional actors (12 female, 12 male), covering eight categorical emotions: neutral, calm, happy, sad, angry, fearful, surprise, and disgust. For evaluation, we employ a speaker-independent 5-fold cross-validation protocol following \cite{luna2021multimodal}. Reported performance is given as the average accuracy over the five folds.

\subsection{Implementation Details}
We trained all models for 50 epochs with a batch size of 4, using the AdamW optimizer.
The learning rate was initialized at $5\times10^{-5}$ and linearly decayed to zero over the course of training.
Our Transformer encoder was configured with $d_{\text{model}}=512$, while projection layers mapped features into a $d_{\text{emb}}=128$-dimensional embedding space.
For the cross-temporal matching loss, we set the Gaussian bandwidth to $\sigma=0.5$, the temperature to $\tau=0.07$, and the weighting coefficient to $\lambda_{\text{ctm}}=0.5$.

\subsection{Comparison with State-of-the-Art Methods}
Table~\ref{tab:main_results} compares audio-visual emotion recognition performance on CREMA-D and RAVDESS. 
Our method achieves 89.49\% on CREMA-D and 89.25\% on RAVDESS, setting new state-of-the-art results on both datasets. 
On CREMA-D, we surpass the previous best method~\cite{lei2023audio} (85.06\%) by 4.43 percentage points, while on RAVDESS, we improve upon ATTSF-Net~\cite{attsfnet_zhang} (88.67\%) by 0.58 percentage points. 

\begin{table}[t]
\centering
\caption{Comparison of audio-visual emotion recognition performance (\%) on CREMA-D and RAVDESS datasets.}%\textbf{Bold} numbers indicate the best accuracy.}
\label{tab:main_results}
\scalebox{0.9}{
\begin{tabular}{l|c|c|c}
\toprule
Method & Year & CREMA-D & RAVDESS \\
\midrule
Ghaleb et al. \cite{ghaleb2019multimodal} & 2019 & 74.00 & 67.70 \\         % rav 10 fold, cre 10 fold
%MSAF \cite{msaf_su} & 2020 & - & 74.86 \\
MATER \cite{mater_ghaleb} & 2020 & 67.20 & 76.30 \\                         % rav 12 fold, cre 10 fold
TA-AVN \cite{taavn_radoi} & 2021 & 84.00 & 78.70 \\                         % rav 8:2, cre 8:2
Luna-Jiménez et al. \cite{luna2021multimodal} & 2021 & - & 80.08 \\
Luna-Jiménez et al. \cite{luna2021proposal} & 2021 & - & 86.70 \\
%Tran et al. \cite{tran2022pre} & 2022 & 70.22 & - \\                        %%% cre 6:2:2
%Guo et al. \cite{guo_conformer}  & 2022 & - & 78.45 \\
%Guo et al. \cite{guo_conformer}  & 2022 & - & 78.45 \\
%Chumachenko et al. & 2022 & 81.58  \\
V8 + A4 \cite{middya2022deep} & 2022 & - & 86.00 \\
%MMCosine \cite{xu2023mmcosine} & 2023 & 66.40 & - \\                        %%% cre 9:1
Mocanu et al. \cite{mocanu2023multimodal} & 2023 & 84.57 & 87.85 \\         %%% rav 10 fold, cre 10 fold
Lei et al. \cite{lei2023audio} & 2023 & 85.06 & - \\                        %%% cre 5 fold
%MA-AVT \cite{mahmud2024ma} & 2024 & 75.20 & - \\                            %%% cre 9:1
%EmoBox \cite{ma2024emobox} & 2024 & 76.75 & 75.32 \\                        %%% rav 6 fold, cre same setting
%ReconBoost \cite{hua2024reconboost} & 2024 & 81.11 & - \\                   %%% cre 9:1
%AGM \cite{li2023boosting} & 2024 & 81.46 & - \\                             %%% cre 8:1:1
%AttA-Net (end-to-end) \cite{attanet_fan} & 2024 & 82.62 \\
AttA-Net \cite{attanet_fan} & 2024 & - & 88.00 \\
HiCMAE \cite{sun2024hicmae} & 2024 & 84.91 & 87.96 \\                       %%% rav 6 fold, cre 5 fold
VQ-MAE-AV \cite{sadok2025vector} & 2025 & 80.40 & 84.80 \\                  %%% rav 6 fold, cre 10 fold
ATTSF-Net \cite{attsfnet_zhang} & 2025 & - & 88.67 \\
\midrule
\textbf{Ours} & 2025 & \textbf{89.49} & \textbf{89.25}\\
\bottomrule
\end{tabular}}
\end{table}

\begin{figure}[t]
\centering
  \includegraphics[width=0.95\linewidth]{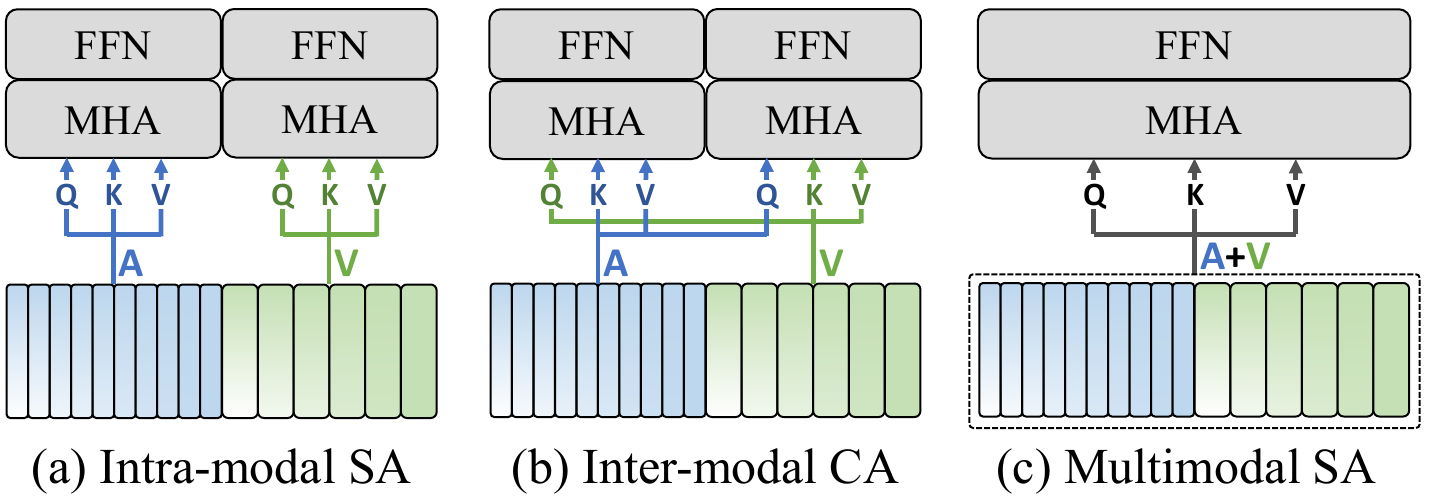}
  \caption{Illustration of differenta fusion strategies using attention mechanisms.}
  \label{fig:transformer}
\end{figure}

\begin{table}[t]
\centering
\caption{Ablation study on different fusion strategies conducted on CREMA-D.}%\textbf{Bold} numbers indicate the best accuracy.}
\vspace{2pt}
\label{tab:ablation_arch}
\begin{tabular}{l|c|c}
\toprule
Fusion Module & $\#$ of parameters & Acc. \\
\midrule
Concat. & - & 85.71 \\
\midrule
ISA + ISA & \multirow{4}{*}{12.61M} & 87.98 \\
ICA + ICA & & 87.49 \\
ISA + ICA & & 87.71 \\
ICA + ISA & & 88.31 \\
\midrule
MSA + MSA (Ours) & 6.83M & \textbf{88.95} \\
\bottomrule
\end{tabular}
\end{table}

\begin{table}[th]
\centering
\caption{Ablation study on positional encoding strategies and CTM loss conducted on CREMA-D.}%\textbf{Bold} numbers indicate the best accuracy.}

\label{tab:ablation_temp_align}
\begin{tabular}{l|c|c}
\toprule
Positional Encoding & w/o $\mathcal{L}_{\text{ctm}}$ & w/ $\mathcal{L}_{\text{ctm}}$ \\
\midrule
Sinusoidal & 88.09 & 88.79 \\
Learnable &  87.44 & 88.79 \\ 
RoPE & 87.76 & 89.00 \\ % 88.10
TaRoPE & 88.95  & \textbf{89.49} \\ % MSA_rope_lr5e5 EMO-689
\bottomrule
\end{tabular}
\end{table}

\subsection{Ablation Studies}
To verify that the unified Transformer encoder can serve as an effective fusion module, we experiment with several design variants illustrated in Fig.~\ref{fig:transformer}.
Prior approaches rely on either intra-modal self-attention (ISA) or inter-modal cross-attention (ICA), whereas our design introduces multimodal self-attention (MSA), which unifies both within a single Transformer block.
For a fair comparison with stacking-based models such as ISA + ICA or ICA + ISA, each variant is configured with two blocks to ensure comparable model capacity.

Table~\ref{tab:ablation_arch} compares different fusion strategies under the common setting of TaRoPE-based positional encoding.
While concatenation provides a lightweight baseline, stacked ISA and ICA variants enhance performance by modeling richer dependencies, but incur greater complexity.
By contrast, our model, which employs two MSA layers, attains the best accuracy with fewer parameters, demonstrating the efficiency of jointly modeling intra- and inter-modal dependencies within a unified block.
Moreover, ISA, ICA, and MSA layers exhibit comparable computational costs, each introducing approximately 0.5 GFLOPs for a 2-second input.

Table~\ref{tab:ablation_temp_align} provides a comprehensive comparison of positional encoding strategies and the contribution of CTM loss.
The proposed TaRoPE surpasses sinusoidal, learnable, and vanilla RoPE embeddings, underscoring the effectiveness of explicit temporal alignment in multimodal fusion.
Moreover, incorporating the CTM loss yields consistent improvements across all variants, confirming its role as a complementary alignment objective.
Overall, the results highlight the importance of temporal alignment as a key design principle for effective multimodal integration.

\subsection{Analysis on Temporal Alignment}
\begin{figure}[t]
\centering
  \includegraphics[width=\linewidth]{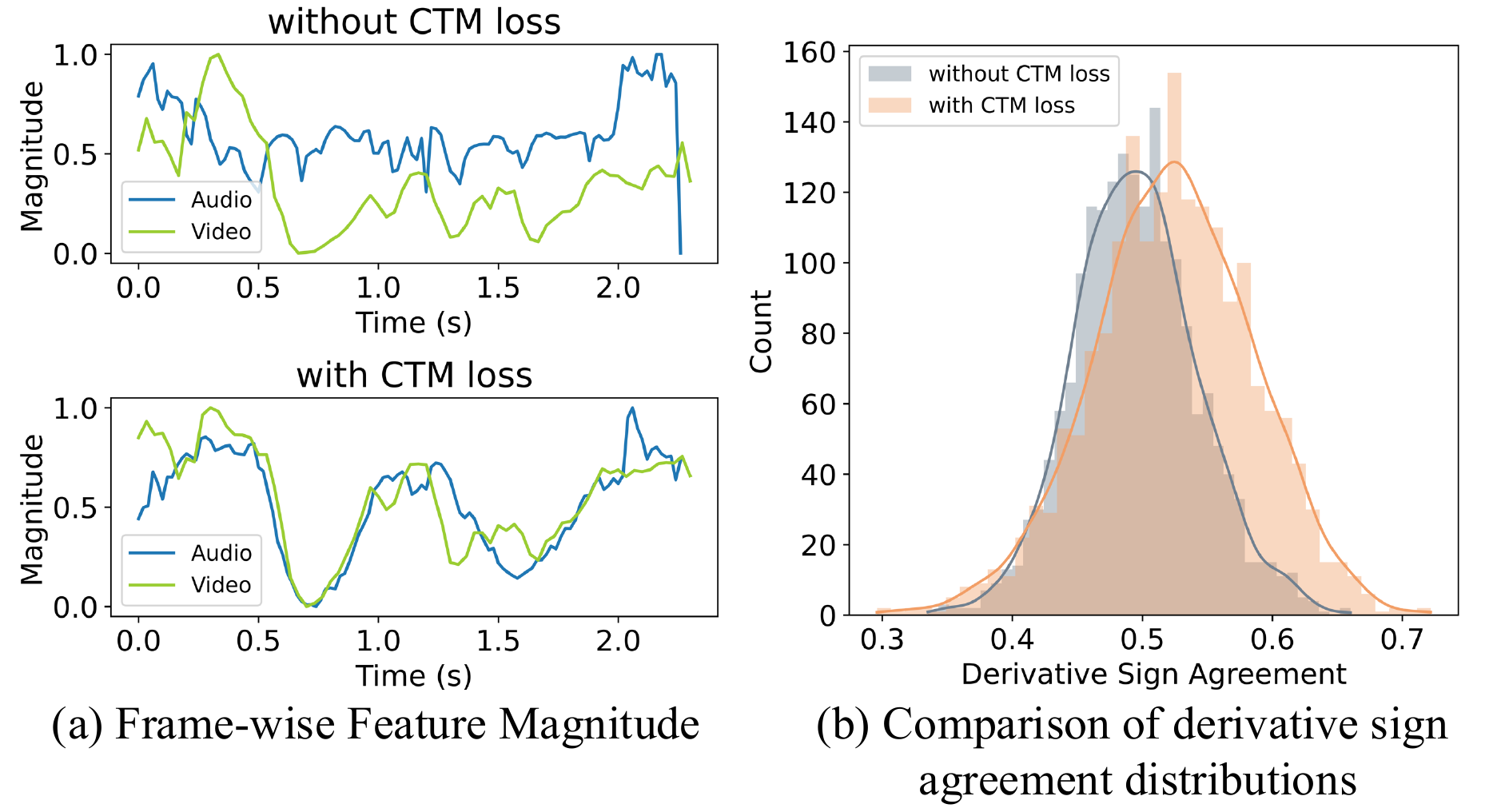}
  \caption{Effect of CTM loss on feature dynamics.}
  \label{fig:analysis}
\end{figure}

We further analyze how the proposed CTM loss influences temporal consistency between modalities. 
Figure~\ref{fig:analysis} illustrates feature dynamics and agreement statistics with and without CTM loss. 

In Fig.~\ref{fig:analysis}~(a), we present an example from the CREMA-D test set (utterance \texttt{1087\_MTI\-ANG\_XX}), where audio and video features are min–max normalized and visualized on the same temporal axis. 
When CTM loss is applied, the audio and video feature magnitudes evolve with similar frame-wise trajectories, suggesting that the auxiliary loss encourages consistent temporal dynamics across modalities.

To ensure this observation is not limited to a single case, 
Fig.~\ref{fig:analysis}~(b) reports derivative sign agreement distributions across the entire test set. 
We find a stronger concentration around high agreement levels when using CTM loss, 
suggesting that temporal trends between audio and video streams are more consistently synchronized at scale. 
These results confirm that CTM loss not only improves recognition accuracy 
but also enhances temporal alignment by enforcing consistency in cross-modal dynamics.

\section{CONCLUSION}
\label{sec:conclusion}

In this paper, we presented a Transformer framework for audio-visual emotion recognition with Temporally-aligned RoPE and a Cross-Temporal Matching loss.
Our framework achieved state-of-the-art performance on CREMA-D and RAVDESS, while ablation studies highlighted the efficiency of the unified architecture and the benefits of temporal alignment.
%These findings suggest promising directions for extending temporally-aware multimodal learning to broader domains.
While this work focuses on controlled benchmarks, extending the proposed framework to large-scale in-the-wild datasets remains an important direction for future work.
\clearpage

% References should be produced using the bibtex program from suitable
% BiBTeX files (here: strings, refs, manuals). The IEEEbib.bst bibliography
% style file from IEEE produces unsorted bibliography list.
% -------------------------------------------------------------------------
\bibliographystyle{IEEEbib}
\bibliography{strings,refs}

@article{lei2023audio,
  title={Audio-visual emotion recognition with preference learning based on intended and multi-modal perceived labels},
  author={Lei, Yuanyuan and Cao, Houwei},
  journal={IEEE Transactions on Affective Computing},
  volume={14},
  number={4},
  pages={2954--2969},
  year={2023},
  publisher={IEEE}
}

@article{mocanu2023multimodal,
  title={Multimodal emotion recognition using cross modal audio-video fusion with attention and deep metric learning},
  author={Mocanu, Bogdan and Tapu, Ruxandra and Zaharia, Titus},
  journal={Image and Vision Computing},
  volume={133},
  pages={104676},
  year={2023},
  publisher={Elsevier}
}

@article{sun2024hicmae,
  title={Hicmae: Hierarchical contrastive masked autoencoder for self-supervised audio-visual emotion recognition},
  author={Sun, Licai and Lian, Zheng and Liu, Bin and Tao, Jianhua},
  journal={Information Fusion},
  volume={108},
  pages={102382},
  year={2024},
  publisher={Elsevier}
}

@article{sadok2025vector,
  title={A vector quantized masked autoencoder for audiovisual speech emotion recognition},
  author={Sadok, Samir and Leglaive, Simon and S{\'e}guier, Renaud},
  journal={Computer Vision and Image Understanding},
  volume={257},
  pages={104362},
  year={2025},
  publisher={Elsevier}
}

@inproceedings{mater_ghaleb,
  title={Multimodal attention-mechanism for temporal emotion recognition},
  author={Ghaleb, Esam and Niehues, Jan and Asteriadis, Stylianos},
  booktitle={2020 IEEE International Conference on Image Processing (ICIP)},
  pages={251--255},
  year={2020},
  organization={IEEE}
}

@inproceedings{ghaleb2019multimodal,
  title={Multimodal and temporal perception of audio-visual cues for emotion recognition},
  author={Ghaleb, Esam and Popa, Mirela and Asteriadis, Stylianos},
  booktitle={2019 8th International Conference on Affective Computing and Intelligent Interaction (ACII)},
  pages={552--558},
  year={2019},
  organization={IEEE}
}

@article{taavn_radoi,
  title={An end-to-end emotion recognition framework based on temporal aggregation of multimodal information},
  author={Radoi, Anamaria and Birhala, Andreea and Ristea, Nicolae-Catalin and Dutu, Liviu-Cristian},
  journal={IEEE Access},
  volume={9},
  pages={135559--135570},
  year={2021},
  publisher={IEEE}
}

@article{luna2021multimodal,
  title={Multimodal emotion recognition on RAVDESS dataset using transfer learning},
  author={Luna-Jim{\'e}nez, Cristina and Griol, David and Callejas, Zoraida and Kleinlein, Ricardo and Montero, Juan M and Fern{\'a}ndez-Mart{\'\i}nez, Fernando},
  journal={Sensors},
  volume={21},
  number={22},
  pages={7665},
  year={2021},
  publisher={MDPI}
}

@article{luna2021proposal,
  title={A proposal for multimodal emotion recognition using aural transformers and action units on ravdess dataset},
  author={Luna-Jim{\'e}nez, Cristina and Kleinlein, Ricardo and Griol, David and Callejas, Zoraida and Montero, Juan M and Fern{\'a}ndez-Mart{\'\i}nez, Fernando},
  journal={Applied Sciences},
  volume={12},
  number={1},
  pages={327},
  year={2021},
  publisher={MDPI}
}

@article{middya2022deep,
  title={Deep learning based multimodal emotion recognition using model-level fusion of audio--visual modalities},
  author={Middya, Asif Iqbal and Nag, Baibhav and Roy, Sarbani},
  journal={Knowledge-based Systems},
  volume={244},
  pages={108580},
  year={2022},
  publisher={Elsevier}
}

@inproceedings{attanet_fan,
  title={AttA-NET: attention aggregation network for audio-visual emotion recognition},
  author={Fan, Ruijia and Liu, Hong and Li, Yidi and Guo, Peini and Wang, Guoquan and Wang, Ti},
  booktitle={IEEE International Conference on Acoustics, Speech and Signal Processing (ICASSP)},
  pages={8030--8034},
  year={2024},
  organization={IEEE}
}

@article{attsfnet_zhang,
  title={ATTSF-Net: Attention-based Similarity Fusion Network for Audio-Visual Emotion Recognition},
  author={Zhang, Jiaming and Zhang, Zhijia and Ju, Zhaojie},
  journal={IEEE Transactions on Affective Computing},
  year={2025},
  publisher={IEEE}
}

@inproceedings{openface_baltruvsaitis,
  author={Baltrusaitis, Tadas and Zadeh, Amir and Lim, Yao Chong and Morency, Louis-Philippe},
  booktitle={2018 13th IEEE International Conference on Automatic Face \& Gesture Recognition (FG 2018)}, 
  title={OpenFace 2.0: Facial Behavior Analysis Toolkit}, 
  year={2018},
  volume={},
  number={},
  pages={59-66},
  doi={10.1109/FG.2018.00019}}

@article{facs_friesen,
  title={Facial action coding system: a technique for the measurement of facial movement},
  author={Friesen, E and Ekman, Paul},
  journal={Palo Alto},
  volume={3},
  number={2},
  pages={5},
  year={1978}
}

@article{xlsr_conneau,
  title={Unsupervised cross-lingual representation learning for speech recognition},
  author={Conneau, Alexis and Baevski, Alexei and Collobert, Ronan and Mohamed, Abdelrahman and Auli, Michael},
  journal={arXiv preprint arXiv:2006.13979},
  year={2020}
}

@article{survey_lian,
  title={A survey of deep learning-based multimodal emotion recognition: Speech, text, and face},
  author={Lian, Hailun and Lu, Cheng and Li, Sunan and Zhao, Yan and Tang, Chuangao and Zong, Yuan},
  journal={Entropy},
  volume={25},
  number={10},
  pages={1440},
  year={2023},
  publisher={MDPI}
}

@article{transformer_vaswani,
  title={Attention is all you need},
  author={Vaswani, Ashish and Shazeer, Noam and Parmar, Niki and Uszkoreit, Jakob and Jones, Llion and Gomez, Aidan N and Kaiser, {\L}ukasz and Polosukhin, Illia},
  journal={Advances in Neural Information Processing Systems (NeurIPS)},
  volume={30},
  year={2017}
}

@article{rope_su,
  title={Roformer: Enhanced transformer with rotary position embedding},
  author={Su, Jianlin and Ahmed, Murtadha and Lu, Yu and Pan, Shengfeng and Bo, Wen and Liu, Yunfeng},
  journal={Neurocomputing},
  volume={568},
  pages={127063},
  year={2024},
  publisher={Elsevier}
}

@article{ravdess_livingstone,
  title={The Ryerson Audio-Visual Database of Emotional Speech and Song (RAVDESS): A dynamic, multimodal set of facial and vocal expressions in North American English},
  author={Livingstone, Steven R and Russo, Frank A},
  journal={PloS one},
  volume={13},
  number={5},
  pages={e0196391},
  year={2018},
  publisher={Public Library of Science San Francisco, CA USA}
}

@inproceedings{emobox_ma,
  title={EmoBox: Multilingual Multi-corpus Speech Emotion Recognition Toolkit and Benchmark},
  author={Ziyang Ma and Mingjie Chen and Hezhao Zhang and Zhisheng Zheng and Wenxi Chen and Xiquan Li and Jiaxin Ye and Xie Chen and Thomas Hain},
  booktitle={Proc. INTERSPEECH},
  year={2024}
}

@article{crema-d_cao,
  title={Crema-d: Crowd-sourced emotional multimodal actors dataset},
  author={Cao, Houwei and Cooper, David G and Keutmann, Michael K and Gur, Ruben C and Nenkova, Ani and Verma, Ragini},
  journal={IEEE Transactions on Affective Computing},
  volume={5},
  number={4},
  pages={377--390},
  year={2014},
  publisher={IEEE}
}

\end{document}